\documentclass[12pt]{article}

\def\bddpi{B \to D_s D^0\pi^0}

\def\be{\begin{equation}}
\def\ee{\end{equation}}
\def\bea{\begin{eqnarray}}
\def\eea{\end{eqnarray}}

\begin{document}

\begin{center}
{\Large \bf Nonstandard CP violation in $\bddpi$ decays}
\vspace*{0.3cm} \\ 
{ \sc A.\ Szynkman}
\vspace*{0.2cm} \\
{IFLP, CONICET $-$ Depto.\ de F\'{\i}sica,
Universidad Nacional de La Plata, \\
C.C. 67, 1900 La Plata, Argentina \\ $\,$\\}
\vspace*{0.5cm}
\begin{abstract}
{\large We study the possibility of measuring nonstandard CP violation effects
through Dalitz plot analysis in \mbox{$\bddpi$} decays. The accuracy in
the extraction of CP violating phases is analyzed by performing a
Monte Carlo simulation of the decays, and the magnitude of possible new
physics effects is discussed. It is found that this represents a hopeful
scenario for the search of new physics.}
\end{abstract}
\end{center}
\vspace*{1.cm}

The origin of CP violation in nature is presently one of the most
important open questions in particle physics. Indeed, the main goal of the experiments devoted to the study of $B$ meson decays is either to confirm the picture offered by the Standard Model (SM) or to provide evidences of CP violation mechanisms originated from new physics. In fact, the common belief is that the SM is nothing but an effective
manifestation of some underlying fundamental theory. In this way, all
tests of the standard mechanism of CP violation, as well as the
exploration of signatures of nonstandard physics, become relevant.

We discuss the possible measurement of nonstandard CP violation in
$\bddpi$~\cite{EGMS}, exploiting the fact that for these processes the asymmetry
between $B^+$ and $B^-$ decays is expected to be negligibly small in the Standard Model. The presence of two resonant channels provides the
necessary interference to allow for CP asymmetries in the differential
decay width, even in the limit of vanishing strong rescattering phases. From the experimental point of view, the usage of charged
$B$ mesons has the advantage of avoiding flavor--tagging difficulties. In addition, the processes $\bddpi$ appear to be statistically favored, in view of their relatively high branching ratios of about 1\%.

In order to measure the CP-odd phases entering the interfering contributions to the total decay amplitude, we propose to use the Dalitz plot fit technique. In general, three body decays of mesons proceed through intermediate resonant channels, and the Dalitz plot fit analysis provides a direct experimental access to the amplitudes and phases of the main contributions. In particular, this fit technique allows a clean disentanglement of relative phases, independent of theoretical uncertainties arising from final state interaction effects. The expected quality of the experimental measurements is estimated by means of a Monte Carlo simulation of the decays, from which we conclude that the phases can be extracted with a statistical error not larger than a couple of degrees, provided that the widths of the intermediate $D^{\ast 0}$ and $D_s^\ast$ resonances are at least of the order of a hundred keV. On the theoretical side, within the framework of generalized factorization we perform a rough estimation of possible nonstandard CP violation effects on the interfering amplitudes. We take as an example the typical case of a multihiggs model, showing that the level of accuracy of the Dalitz plot fit measurements can be sufficient to reveal effects of new physics.

Let us finally stress that tree-dominated decays like $\bddpi$ are usually not regarded as good candidates to reveal new physics, since the effects on branching ratios are not expected to be strong enough to be separated from the theoretical errors. Our proposal represents a possible way of detecting these effects by means of CP asymmetries, which can allow the disentanglement of new physics contributions to penguin-like operators in a theoretically simple way.

\section*{Acknowledgements}

A.S. acknowledges financial aid from Fundaci\'on Antorchas (Argentina).

\end{document}